\documentclass[aps,prl,twocolumn]{revtex4}
\usepackage{times}
\usepackage{latexsym}
\usepackage{graphicx}
\usepackage{verbatim,times,bbm}
\usepackage{color}
\usepackage{appendix}
\usepackage{bm}

\def\uno{\mbox{1 \kern-.59em {\rm l}}}
\def\th{\theta}
\def\ph{\phi}
\def\be{\begin{equation}}
\def\ee{\end{equation}}
\def\ba{\begin{eqnarray}}
\def\ea{\end{eqnarray}}
\def\lo{\longrightarrow}

\def\la{\langle}
\def\ra{\rangle}

\def\n{\nu}

\def\n{\hat{{\bf n}}}

\begin{document}

\title{Scaling of macroscopic superpositions close to a quantum phase transition}

\author{Tahereh Abad, and Vahid Karimipour}
\affiliation{ Department of Physics, Sharif University of Technology, Tehran, Iran}

\begin{abstract}
It is well known that in a quantum phase transition (QPT), entanglement remains short ranged [Osterloh et al., Nature 416 608-610 (2005)]. We ask if there is a quantum property entailing the whole system which diverges near this point. Using the recently proposed measures of quantum macroscopisity,  we show that near a quantum critical point, it is the effective size of macroscopic superposition between the two symmetry breaking states which grows to the scale of system size and its derivative with respect to the coupling shows both singular behavior and scaling properties.  \end{abstract}


\maketitle

{\bf Introduction}
Superposition of states is the most important and distinctive feature of the microscopic world:   atoms, electrons, and photons, and even molecules \cite{zeilinger, arndt}  can exist in a superposition of two or more physical states. This property lies at the heart of all intriguing quantum phenomena that we know, from wave-particle duality to entanglement and non-locality. Nevertheless, everyday objects of macroscopic size do not exist in superposition of their different states. We do not see states which represent a cat in a superposition of dead and alive states \cite{Schrodinger}. This maybe due to the fact that quantum mechanics is modified at a certain scale \cite{GRW, BassiReview} or it maybe due to the extremely rapid decoherence of such states as the result of their macroscopic number of constituents with the outside world \cite{zurek}. There is now intensive experimental \cite{zeilinger, arndt} and theoretical investigations \cite{BassiReview} to reveal if there is a distinct border between the micro and macro world.\\

Despite the absence of quantum superpositions on macroscopic scales, we see many macroscopic phenomena which are the result of collective quantum mechanical behavior of their microscopic constituents. An important example is the phenomenon of superconductivity which arises  when pairs of electrons interact with phonons and tend to entangle and  form Cooper pairs \cite{BCS}, which subsequently undergo Bose condensation. Without this microscopic superposition of atomic entities, we do not have the phenomenon of superconductivity at the macroscopic level. Nevertheless this cannot be called a witness of quantum mechanics at macroscopic scale. It was Leggett \cite{Leggett} who first emphasized the difference between this type of macroscopic quantum effect  which is the result of collective superposition of microscopic entities and the one which was discussed above, i.e. when a macroscopic system, a large macromolecule \cite{zeilinger}, a virus or a cat \cite{Schrodinger}, is in a superposition of two macroscopically distinct states. \\

To put all this in very concrete terms, one can take a number $N\sim {\rm Avogadro \ number}$ of spin 1/2 particles. Then the simplest conceivable example of the former is a state like $\left[\frac{1}{\sqrt{2}}(|\uparrow\ra + |\downarrow\ra)\right]^{\otimes N}$ and of the later a Greenberger-Horne-Zeilinger state \cite{GHZ}  $|GHZ\ra=\frac{1}{\sqrt{2}}(|\uparrow\ra^{\otimes N}+|\downarrow\ra^{\otimes N})$. It is only the later kind that is called a quantum superposition on a genuinely macroscopic scale. We should stress on the difference between three concepts,  macroscopic quantum superposition, entanglement and quantum correlation. By macroscopic superposition we mean a state like $|\psi_1\ra+|\psi_2\ra$, where $|\psi_1\ra$ and $|\psi_2\ra$ are macroscopically distinct \cite{Korsbakken}, in which case a  measurement of any single particle can reveal the difference between $|\psi_1\ra\equiv |\uparrow\ra^{\otimes N}$ and $|\psi_2\ra\equiv |\downarrow\ra^{\otimes N}$. States with this property are a priori extremely rare in many ensembles of pure states, especially there are highly entangled states that feature vanishing macroscopic superposition as pointed out in \cite{Molmer}. Finally this property is distinct from  quantum correlation which can exist even in separable mixed states and is measured by quantum discord \cite{Ollivier}.

It is quite conceivable that there is a spectrum of states between these two extremes. Therefore, like many other quantities, once the qualitative difference and the typical examples of these two classes of states are understood, the next step is to define reasonable measures to quantify exactly how macroscopic a quantum superposition is. These measures are called measures of quantum macrosopicity. To this end, various proposals have been suggested  \cite{Dur, Shimizu and Miyadera, Bjork, Shimizu, Korsbakken, Marquardt, Lee, Frowis, Yadin}. The basic idea used in all these proposals is that macroscopic superposition entails a large amount of uncertainty when a suitably chosen macroscopic observable is measured. The way this large variance scales with $N$ defines a measure of macroscopisity. \\

We now come to the main questions asked in this letter: Is there macroscopic superposition of symmetry breaking states in a quantum phase transition (QPT), and if yes, how this macroscopicity behaves near the critical point? How the relevant critical exponents are related with this scaling?  What happens exactly near a point of quantum phase transition? The intuition behind these questions is the observation that in a classical phase transition, say in the Ising model, large areas of up and down spins co-exist. So one expects that in a quantum phase transition, this co-existence appears in the form of large scale superpositions. Therefore an exponent like $\nu$ which sets the scale of divergence of correlation length in classical phase transition, may appear here in some form of  divergence of macroscopic superposition. \\

In fact, it was anticipated that near a quantum critical point, we will see long range entanglement (a quantum parallel of diverging correlation length in classical phase transition). However it came as a big surprise \cite{Osterloh} that near a critical point, entanglement always remain short-ranged. Instead what becomes singular and shows scaling behavior is  the derivative of this short range entanglement. Later elaborations revealed a more detailed picture. First it was shown that  while entanglement remains short range in a QPT, other kinds of quantum correlations, measured by discord, can be long ranged \cite{Mazieroa}. Then it was shown that depending on whether the QPT is first order or continuous, the entanglement itself or its derivative is the quantity which is relevant in detecting it \cite{Wu1, Wu2}. \\

Therefore this  result, gives some weight to the question we are asking. In fact by using the measures of macroscopic  superposition which have been recently developed \cite{Shimizu and Miyadera, Frowis}, we show that near the point of phase transition, the system, which is undecided between the two symmetry breaking phases,  actually goes into a macroscopic superposition of them, and shows singular behavior and scaling properties in the derivative of the effective size of this superposition.  \\

Our findings on macroscopic superposition, in conjunction with other results on short-range entanglement in quantum phase transitions \cite{Osborne, Vidal, Osterloh, Hofmann, Zhu, Cozzini, Alipour}, confirm the recent results of \cite{Molmer} according to which, 
macroscopicity is rather rare in ensembles of random pure states despite having large geometric entanglement. \\

{\bf The Ising model in transverse field} 
We now consider a paradigmatic example of quantum phase transitions \cite{Sachdev}, namely the Ising model in transverse field, described by the Hamiltonian
\be\label{IsingHam}
H=- \lambda \sum^{N}_{i=1} \sigma^{x}_{i} \sigma^{x}_{i+1}-\sum^{N}_{i=1} \sigma^{z}_{i},
\end{equation}
where $\sigma^{\mu}_{i}$ is the $\mu-$th Pauli matrix $(\mu = x, y, z)$ at site $i$ and $\lambda$ is the inverse strength of
the external field. Besides the obvious translational symmetry,  the Hamiltonian is real and has a global phase flip symmetry $[U, H]=0$, where 
$ U=\prod^{N}_{i=1} \sigma^{z}_{i}.$ The operator $U$ flips $\sigma^x_i$ (and also $\sigma^y_i$) and leaves $\sigma^z_i$ unchanged. Let the state vectors $\{|0\ra, |1\ra\}$ and $\{|+\ra, |-\ra\}$ be the eigenstates of the $\sigma^{z}$ and $\sigma^{x}$, respectively. In the limit $\lambda=0$ the unique ground state $|00\cdots 0\ra$ is unchanged under the $U$ symmetry, while in the limit $\lambda\lo \infty$ the symmetry is broken (in the thermodynamic limit) and the two degenerate ground states $ |+ +  \cdots +  \rangle\ $ and $|- -\cdots -   \rangle$ are mapped to each other by $U$. The model is solved exactly by using the Jordan-Wigner transformation which turns it into a free fermion model \cite{Lieb, Pfeuty}. At zero temperature and in the limit $\lambda=0$, the system goes to the unique ground state $|00\cdots 0\ra$ where all the spins are aligned in the $z-$direction. In this phase $\la \sigma^x\ra$ as a local order parameter vanishes. As we gradually increase the value of $\lambda$,  at $\lambda=1$, the system undergoes a quantum phase transition, the symmetry break downs and the system chooses one of the two degenerate ground states, finally in the limit $\lambda\lo \infty$ system goes to $|++\cdots +\ra$ or $|--\cdots -\ra$, where each of these two states shows superpositions (in the $\{|0\ra, |1\ra\}$ basis), but on a microscopic scale of each individual spin. \\

{\bf Measures of macroscopic superposition} 

We now remind the reader of a few basic facts about two measures of macroscopicity which we mainly use in our analysis. A simple calculation shows that the variance of any additive operator like ${\cal A}=\sum_{i=1}^N A_i$ on any product state $|\Psi\ra=|\psi\ra^{\otimes N}$ is proportional to $N$, that is 
${\cal V}_\Psi( {\cal A})=N{\cal V}_{\psi}(A)$, where ${\cal V}_\phi(X):=\la \phi|X^2|\phi\ra-\la \phi|X|\phi\ra^2$ is the variance of the observable $X$ on the state $|\phi\ra.$
 However quantum states which are in macroscopic superpositions show quadratic behavior when the variance of suitable additive operators are measured on them. An example is the GHZ state. For this state the observable $M_z:=\sum^N_{i=1} \sigma^z_i$ (e.g. the magnetization in the $z-$direction) shows an anomalously large variance. In fact straightforward calculation shows that  ${\cal V}_{GHZ}(M_z):=\langle M_z^2\rangle-\langle M_z \rangle^2=N^2$. These anomalous large  fluctuations are the signature of a macroscopic quantum superposition. In view of the fact that superposition is the characteristic feature of quantum mechanics against classical mechanics, it is usually said that the scaling of an additive operator with system size $N$,  is taken to depict its classical or quantum behavior. However it should be noted that this quantum versus classical division is to be interpreted as superposition versus product state. There are other divisions with respect to entanglement or more generally the type of correlations \cite{LiLuo} which, although have their root in the superposition property are defined and characterized in a different way. Here we confine ourselves to this specific meaning mentioned above.  Based on this concept, the $p-$ index of a pure state  $|\psi\rangle$ is defined as \cite{Shimizu and Miyadera}:
 
 \begin{equation}
\max_{A\in {\cal A}}{\cal V}_{\psi}(A)={\cal O}(N^{p}),\ \  N\ \ {\rm large},
\end{equation}
where ${\cal A}$ is the set of all additive operators $A=\sum^N_{i=1} A_i$ such that every operator $A_i$ acts non-trivially on $i-$th particle and $||A_i||=1$. A fully product state has $p=1$. This means that $p > 1$ is an entanglement witness for pure states. The state with $p = 2$ contains superposition of macroscopically distinct states, because in this case a Hermitian additive operator has a "macroscopically large" fluctuation in the sense that the relative fluctuation does not vanish in the thermodynamic limit.

For such pure states, the fluctuation of an observable means the existence of a superposition of eigenvectors of that observable corresponding to different eigenvalues with the largest difference. An example is the $GHZ$ state which is the superposition of two states which are eigenvectors of the additive operator $\sum_{i=1}^N \sigma_z$ corresponding to $N$ and $-N$.  \\

As another measure of macroscopisity,  Fr\"{o}wis and D\"{u}r defined the quantum fisher information \cite{Frowis} which detects a certain kind of correlation. It is well known that a separable state has a Fisher information  which scales at most linearly with the system size $N$ for every local operator like $A$ \cite{Pezze}, ${\cal F}(\rho_{sep}, A)\leq4N$. ( For the definition of Fisher information and its properties see \cite{Paris}.) On the other hand, for $GHZ$, ${\cal F}(GHZ, A)=4N^{2}$. The authors of \cite{Frowis} introduce the concept of an "effective size" $N_{\text{eff}}$.  For a general state $\rho$ of $N$ particles, this measure is defined as
\begin{equation}
N_{\text{eff}}(\rho):=\max_{A\in {\cal A}} {\cal F}(\rho, A)/(4N).
\end{equation}
In other words, $N_{\text{eff}}$ defines the scale over which macroscopic superposition and hence quantum behavior prevails.  If $N_{\text{eff}}(\rho)={\cal O}(N)$, we have macroscopic quantum behavior while if $N_{\text{eff}}(\rho)={\cal O}(1)$, then quantum behavior, if existing at all, exists at the microscopic level. For pure states $\rho=|\psi\rangle \langle\psi|$, the Fisher information reduces to $4$ times of the variance ${\cal V}_{\psi}(A)=\langle \psi|A^2|\psi\rangle-\langle \psi|A|\psi\rangle^2$, and the effective size takes the form
 \begin{equation}\label{effSize}
N_{\text{eff}}(\psi)=\max_{A\in {\cal A}} {\cal V}_{\psi}(A)/N.
 \end{equation}
In the sequel we will use these two measures to quantify the macroscopic superposition of the ground state of the Ising model in transverse field, when it  undergoes a quantum phase transition. \\

{\bf Macroscopic superposition and its scaling behavior }
Let us denote the ground state of the Transverse Ising Model (\ref{IsingHam}), by $|\psi\ra$, which simplifies the more detailed notation $|\psi_\lambda(N)\ra$.  Let $A$ be of the form  $A_{\bf n}=\sum^N_{i=1}\bm{\sigma}_{i}\cdot {\bf n}$, where ${\bf n}=(\sin \th \cos \ph, \sin \th \sin \ph, \cos \th)$. In order to use measure of macroscopicity (\ref{effSize}),  we have to first determine the direction ${\bf n}$ for which maximum variance is obtained. 
Using the definition of variance we find 
\begin{eqnarray} \label{expandVar} \nonumber
{\cal V}_{\psi}(A_{\bf n})&=&\sin^{2}\theta \left[\cos^{2}\phi\  \langle X^{2}\rangle+\sin^{2}\phi \ \langle Y^{2} \rangle\right] +\cos^{2}\theta\langle Z^{2}\rangle\\ \nonumber
&+&\frac{1}{2}\sin 2\theta \left[\cos \phi\ \langle ZX+XZ \rangle+\sin \phi \ \langle ZY+YZ \rangle\right] \\ \nonumber &+&\frac{1}{2}\sin^{2} \theta \sin 2\phi\  \langle XY+YX \rangle \\ 
&-&\langle X\sin \theta  \cos \phi +Y\sin \theta  \sin \phi +Z \cos \theta  \rangle^{2},
\end{eqnarray}
where $X=\sum^N_{i=1} \sigma^{x}_{i}$ with similar definitions for $Y$ and $Z$.  The $U$ symmetry (by which $(X,Y,Z)\lo (-X,-Y,Z)$), implies that  $\la Y\ra=\la X\ra=\la XZ+ZX\ra=\la YZ+ZY\ra=0$. As for $\la XY+YX\ra$, it vanishes due to the reality of the Hamiltonian and Hermiticity of the operator $XY+YX$ and the fact that $Y^*=-Y$. Therefore the variance (\ref{expandVar}) is reduced to
\begin{eqnarray} \label{}
{\cal V}_{\psi}(A_{\bf n})&=&\sin^{2}\theta\left[ \langle X^{2}\rangle \ \cos^{2} \phi +\langle Y^{2}\  \rangle \sin^{2}\phi\right] \\ \nonumber &+&\cos^{2}\theta\left[\langle Z^{2}\rangle-\langle Z\rangle^{2}\right].
\end{eqnarray}
To find the direction ${\bf n}$ for which this variance is maximized, we note that since ${\bf n}$ is defined on the compact surface of a 2-sphere, the maximum will be a local one which is determined by examining the first and second  derivatives of ${\cal V}_{\psi}(A_{\bf n})$ as a function of $\theta$ and $\phi$. For simplicity let us denote this quantity simply by ${\cal V}$. It is then found that 

\begin{eqnarray}\label{diffphi}\nonumber
\frac{\partial {\cal V}}{\partial \theta}&=&\sin 2\theta \left[\cos^{2} \phi \langle X^{2}\rangle+\sin^{2} \phi \langle Y^{2}\rangle-\langle Z^{2}\rangle+\langle Z\rangle^{2}\right],\\
\frac{\partial {\cal V}}{\partial \phi}&=&\sin^{2}\theta\ \sin 2\phi \left[\langle Y^{2}\rangle-\langle X^{2}\rangle\right].\label{difftheta}
\end{eqnarray}
The extrema are obtained by setting these two derivatives  equal to zero which yield three solutions $(\theta=0,\ \phi = {\rm irrelevant})$ or ${\bf n}={\bf z}$,   $(\theta=\frac{\pi}{2}, \phi=0)$ or ${\bf n}={\bf x}$ and  $(\theta = \frac{\pi}{2}, \phi=\frac{\pi}{2})$ or ${\bf n}={\bf y}$.  Examining the Hessian matrix (of second derivatives) shows that ${\bf n}={\bf x}$ is indeed the point of maximum. Note that while the model has a U symmetry, it lacks rotational symmetry and hence in general $\la X^2\ra \ne \la Y^2\ra$. In other words, the $U$ operator cannot change an operators $X$ into $Y$ or vice versa. Therefore the observable which detects macroscopic superposition is $A_{\bf x}=\sum_{i=1}^N \bm{\sigma}_i\cdot{\bf x}$, which from (\ref{effSize}) gives the effective size of superpositions as 
\be\label{effFinal}
N_{\text{eff}}\equiv \frac{{\cal V}_{\psi}(A_{\bf x})}{N}=\frac{\la X^2\ra}{N}=\sum_{n=1}^N\la \sigma^x_1\sigma^x_{n}\ra,
\ee
where in the last equality we have used translational invariance of the system. Therefore determination of the macroscopic measure of superposition is reduced to calculation of the two point functions $G^{xx}(n):=\la \sigma^x_1\sigma^x_{1+n}\ra$.  Note that while the model is a free fermion, determination of the two-point spin functions is quite non-trivial due to the non-local nature of the Jordan-Wigner transformation.  These two-point functions have been determined in \cite{Lieb, Pfeuty}. They are given by
$$
G^{xx}(n):=\left |
        \begin{array}{cccccccc}
          G_{-1} & G_{-2} & . . . & G_{-n}\\
          G_{0} & G_{-1} & . . . & G_{-n+1}\\
          . & . & &.\\
             . & . & &.\\
                . & . & &.\\
                 G_{n-2} & G_{n-1} & . . . & G_{-1}\\
         
        \end{array}
      \right |,
$$ 
with
$
G_n=L_n+\lambda L_{n+1},
$
and
$
L_n=\frac{2}{N}\sum_{k>0} \lambda^{-1}_k \cos(kn),
$
where
$
\Lambda_k=\sqrt{1+\lambda^2+ 2\lambda \cos(k)}, \ \ \  m=0, 1, ..., \frac{1}{2}(N-1).
$

We now want to use (\ref{effFinal}) and  see how this effective size changes as we change the coupling $\lambda$. In the two limiting case, the behavior is simple and expected. When $\lambda=0$, the ground state is $|\psi\ra_{\lambda=0}=|00\cdots 0\ra$, for which $\la \sigma^x_1\sigma^x_n\ra_{n\ne 1}=0$. Therefore we obviously have $N_{\text{eff}}=\la \sigma^x_1\sigma^x_{1}\ra=1$ and hence $N_{\text{eff}}=1$, implying no superposition. On the other hand in the limit of very large $\lambda$, the first term of the Hamiltonian (\ref{IsingHam}) dominates and the ground state goes to $|\psi\ra_{\lambda\lo \infty}\approx \frac{1}{\sqrt{2}}(|++\cdots + \ra + |- - \cdots - \ra)$, where it is expected that $N_{\text{eff}}\lo N$. This is indeed the case as a simple calculation from (\ref{effFinal}) shows. \\

{\bf Remark} Note that in the limit $\lambda\lo \infty$, as long as $N$ is finite, no symmetry breaking happens and so the ground state in the limit does not break the $U$ symmetry. So starting at $\lambda=0$ with the state $|00\cdots 0\ra$ which is an eigenstate of $U$ with eigenvalue $1$, when we continuously change $\lambda$, we always remain in the same eigenspace and hence in the limit $\lambda\lo \infty$, we have the superposition$|\psi\ra_{  \lambda\lo \infty}\approx \frac{1}{\sqrt{2}}(|++\cdots + \ra + |- - \cdots - \ra)$ and not one of its individual components. It is only in the thermodynamic limit that the symmetry breaks down and only one of the two states is chosen and hence the macroscopic measure again gives a zero value as $\lambda=0$.\\

In the absence of symmetry breaking, which only happens for an infinite system, we have to use (\ref{effFinal}) and do a finite size scaling.  The results are shown in figure (1).  It is clear from this figure that $\frac{N_{\text{eff}}}{N}$ sharply goes from $0$ to $1$ at a point $\lambda_m(N)$. As $N$ approaches $\infty$, this transition becomes discontinuous. The inset of figure (1) expresses this in an alternative way by showing the $p-$index as a function of $\lambda$ where again a sharp transition is found at $\lambda_m(N)$. Therefore the quantum phase transition is concomitant with a divergence of the derivative of $N_{\text{eff}}$ which sets the scale of macrosocopic superpositions in the system, that is, near a critical point, quantum superposition of the two degenerate states entails the whole system. Obviously this divergence is not seen directly for finite N. Instead what we see is that $\frac{dN_{\text{eff}}}{d\lambda}$ is a rapidly  increasing function of N. Let us see the position of the maximum, $\lambda_m(N)$ and the maximum value of the derivative behave with N. Figure (2) shows that by increasing $N$ toward the thermodynamic limit, the position of this transition point approaches the actual critical value $\lambda_c:=\lambda_m(\infty)=1$. This approach toward the actual critical point is governed by a power law in the form  
\be\label{positionmaximum}
1-\lambda_m(N)\sim N^{-1.96}.
\ee
We note in passing that the same scaling behavior, albeit with different powers, has been reported in other works. In particular in \cite{Osterloh} where the nearest neighbor concurrence or entanglemnt is considered the relation is $1-\lambda_m(N)\sim N^{-1.87}$ and in \cite{Hofmann} where negativity of three consecutive particles is considered the relation is like $1-\lambda_m(N)\sim N^{-2.19}$. Finally in \cite{Zhu} where the maximum of geometric phase of the ground state is considered the approach toward the actual critical point is like $1-\lambda_m(N)\sim N^{-1.803}$. 

\begin{figure}[t]\label{figVar}
\centering
\includegraphics[width=8cm,height=5.5cm,angle=0]{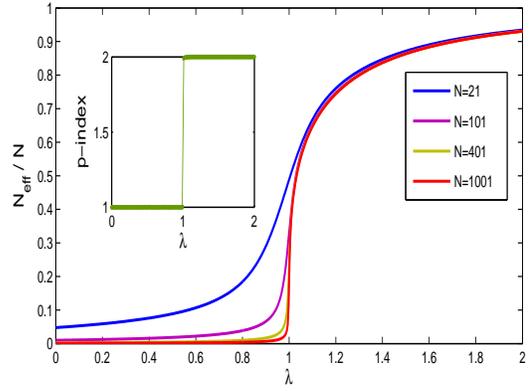}
\caption{The macroscopic superposition as measured by effective size $N_{\text{eff}}:=\frac{\la X^2\ra}{N}$ in the transverse Ising model as a function of the coupling $\lambda$. The curves from top to bottom, correspond to $N=21, 101, 401$ and $N=1001$. The inset shows the $p-$index as a function of $\lambda$. We considered different system sizes,  $1000<N<2000$,  for which a sharp transition in the $p-$ index could be detected. It is seen that both $N_{\text{eff}}$ and $p$ change discontinuously near the point $\lambda_c=1$. }
\end{figure}

Furthermore, the maximum value of the derivative at $\lambda_m(N)$ diverges with $N$ as follows 
\be\label{valuemaximum}
\frac{dN_{\text{eff}}}{d\lambda}(\lambda_m, N)\sim N^{1.75},
\ee
leading to a divergent behavior in the thermodynamic limit $(N\lo \infty)$.
\begin{figure}[t]\label{figDiffVar}
\centering
\includegraphics[width=8cm,height=5.5cm,angle=0]{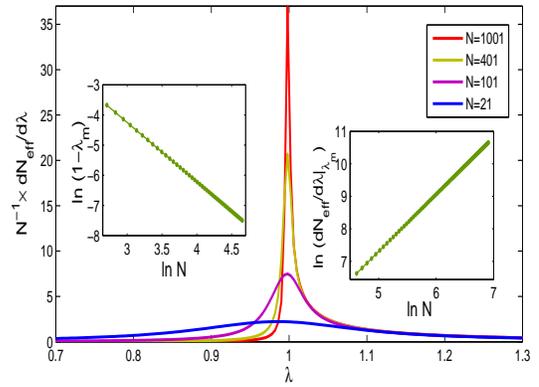}
\caption{The derivative of macroscopicity as measured by $d N_{\text{eff}}/d \lambda$, obtains its  maximum value at point $\lambda_m(N)$. The left inset shows that this point approaches the actual critical point 
$\lambda_c:=\lambda_m(\infty)=1$ as in (\ref{positionmaximum}) and the right inset shows that the maximum value itself diverges with N as in (\ref{valuemaximum}). 
}
\end{figure}
Finally the most important exponent, namely $\nu=1$ which puts the model in the universality class of the Ising model is obtained when the data for the function $\frac{dN_{\text{eff}}}{d\lambda}(\lambda, N)$ collapse to a single curve by choosing a suitable scaling function. Figure (3) shows such a scaling function. It is clearly seen that   
\be\label{scalingfunction}\frac{dN_{\text{eff}}}{d\lambda}(\lambda, N)-\frac{dN_{\text{eff}}}{d\lambda}(\lambda_m, N)=N^{1.89}Q(N(\lambda-\lambda_m)),\ee
 where $Q$ is a universal function derived numerially and shown in figure (3). When this is compared with the scaling function with variable $N^{1/\nu}(\lambda-\lambda_m)$, \cite{Barber}, it leads to $\nu=1$, in agreement with  the well knownresults for the Ising model in transverse field \cite{Lieb, Pfeuty, Barouch}. On the other hand, we can find the asymptotic behavior of $\frac{dN^*_{\text{eff}}}{d\lambda}(\lambda):=\frac{dN_{\text{eff}}}{d\lambda}(\lambda, N\rightarrow \infty)$ in terms of the coupling $\lambda_c-\lambda=1-\lambda$. For this we have to take very large values of $N$ to mimic the thermodynamic limit $N\lo \infty$. For $N=4001$, the result is 
\be\label{kappa}
\frac{dN^*_{\text{eff}}}{d\lambda}(\lambda)\sim (1-\lambda)^{-1.89},
\ee
as we see the exponent $1.89$ is fully consistent with the equation (\ref{scalingfunction}).\\
 
Note that in the light of (\ref{scalingfunction}),  equation (\ref{valuemaximum}) is consistent with a recent result of \cite{Philipp} on measuring multipartite entanglement by dynamical susceptibilities,  where it is reported that $f_Q:=\frac{F_Q}{N}\sim N^{0.75}$. Here $F_Q$ is the Fisher information which for pure states reduces to four times the variance. The consistency comes about by noting that $f_Q=4N_{\text{eff}}$ and a relation like $f_Q\sim N^{0.75} h((\lambda-\lambda_m)N)$, leads to the relation $\frac{df_Q}{d\lambda}\equiv \frac{dN_{\text{eff}}}{d\lambda}\sim N^{1.75}h'((\lambda-\lambda_m)N)$, where $h'$ is the derivative of $h$. \\

Finally It is instructive to study a hypothetical transition between the ground state at $ \lambda=0$ ($|0\ra^{\otimes N}$) and the ground state at $\lambda \lo\infty $ ($\frac{1}{\sqrt{2}}\left[|+\ra^{\otimes N}+|-\ra^{\otimes N}\right]$), in the form of 
\be
|\psi_n\ra=\frac{1}{\sqrt{2}}\left[|+\ra^{\otimes n}+|-\ra^{\otimes n}\right]\otimes |0\ra^{N-n},
\ee
where a domain wall has been created at site $n$. The state is unzipped at point $n$ and as $n$ grows, the hypothetical state changes from the ground state at $\lambda=0$ to the ground state at $\lambda\lo \infty$. It is a simple calculation to calculate $N_{\text{eff}}$ for this state. It turns out to be 
\be
N_{\text{eff}}=\frac{n(n-1)}{N}+1,
\ee
which shows that the effective size raises from $1$ to $N$ in a smooth way. \\
\begin{figure}[t]\label{figScaling}
\centering
\includegraphics[width=8cm,height=5.5cm,angle=0]{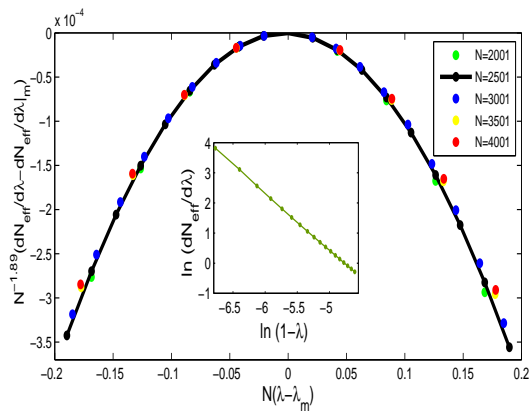}
\caption{The collapse of the derivative of macroscopicity on a universal function of $N(\lambda-\lambda_m)$, equation (\ref{scalingfunction}), which shows that $\n=1$. The inset shows the asymptotic behavior of $d N_{\text{eff}}/d \lambda$ in the vicinity of the quantum criticality as in equation (\ref{kappa}), the result is for $N=4001$. }
\end{figure}

In summary, by studying a paradigmatic example of quantum phase transitions and using the recently defined measures of macroscopic superpositions we have shown that the degree of macroscopic superposition diverges near the point of quantum phase transitions. Physically this means that quantum phase transitions and the ensuing symmetry breaking does not happen on a microscopic scale which then would  propagate through the whole system, but the entire  system goes into a macroscopic superposition of symmetry breaking states. This change happens in a very sharp way which becomes discontinuous in the thermodynamic limit. These findings verify, by different tools,  the scenario suggested in \cite{zurek2} according to which, "a topological defect can be put in a non- local superposition, where the order parameter of the system is “undecided” by being in a quantum superposition of conflicting choices of the broken symmetry."  The final stage of quantum phase transition is achieved by the rapid environment-induced decoherence ($\tau_{dec}\sim \frac{1}{N}$) of this macroscopic superposition to a statistical mixture of the two symmetry breaking phases. Such statistical mixtures also show a non-analytical behavior, measured by the quantum Fisher information, but they do not show diverging behavior \cite{Jin}. These findings on macroscopic superposition, in conjunction with the result of \cite{Osterloh} on short-ranged entanglement, may be an example of the interplay between the two properties, recently mentioned in \cite{Molmer}. 

{}

\end{document}